# Design and Implementation of a Blade-Type Linear Paul Trap

L. E. Filgueira,[1, 2] M. Luda,[1, 3] and C. T. Schmiegelow[1, 2]

[1])*Universidad de Buenos Aires, Facultad de Ciencias Exactas y Naturales, Departamento de Física. Buenos Aires, Argentina.*

[2])*CONICET - Universidad de Buenos Aires, Instituto de Física de Buenos Aires (IFIBA). Buenos Aires, Argentina.*

[3])*CEILAP, CITEDEF, J.B. de La Salle 4397, 1603 Villa Martelli, Buenos Aires, Argentina*

(Dated: 21 April 2025)

In this work, we present the design, implementation, and construction of a linear Paul trap in a blade configuration. The trap was optimized to minimize micromotion and enable the formation of linear ion chains comprising tens of ions. We use ytterbium atoms, and in particular, we describe the trapping and cooling of the isotopes $^{174}$Yb$^+$ and $^{171}$Yb$^+$, starting from an isotope-selective ionization process based on laser frequency stabilization and optimization of fluorescence for each isotope. The electronic control system—including magnetic field generation, laser delivery, and microwave driving—was fully implemented and is described in detail. The system supports pulsed operation, and we performed Rabi oscillations of the hyperfine states in the $^2S_{1/2}$ level, demonstrating coherent spectroscopy. These capabilities are fundamental tools for the experiments we aim to pursue, including the use of structured light beams, quantum simulation protocols, and the generation of non-classical motional states.

## I. INTRODUCTION

Trapped ions are a versatile experimental platform used in diverse areas such as precision metrology[1,2], the exploration of fundamental physics[3,4], and applications in quantum simulation[5,6] and quantum computation[7,8]. Each of these areas comes with specific requirements, which have driven the development and optimization of trap geometries, construction techniques, and control systems in various directions. For example, ion traps optimized for single ions have been employed in frequency metrology[9,10] or for interfacing atoms with fiber optics and light collection systems[11–15]. Structured linear traps that enable splitting and merging of ion crystals have been developed for quantum computation within the quantum CCD architecture[16–20]. Bulk linear traps, in turn, allow robust confinement of ions and have found use in both quantum information and fundamental physics experiments[3,13,21,22].

In our case, we aim for a trap design that offers generous optical access and supports the stable operation of linear ion chains. Our objective was to design and construct a platform suitable for quantum optics experiments with structured light[23–25], precision frequency metrology, and the manipulation of long ion chains. This enables the study of both collective behavior in many-body systems and the creation and characterization of non-classical motional states. To meet these goals, we opted for a linear Paul trap with a blade-type electrode configuration[26,27]. The trap was designed and built with a focus on mechanical robustness, high optical accessibility, low capacitance, and minimal axial micromotion.

The following sections are organized following the development of the trap. In the first section a brief theoretical depiction of the linear trap is carried followed by the main design considerations which includes the simulation, the construction and assembly and another important factors like vacuum system and the electric potentials of the trap. The second section is about the operation of the trap, where the detection of fluorescence is presented, along with the first characterization for the use of the trap like the ionization of neutral ytterbium, the level structure of the isotopes used in the trap and the main components like lasers, magnetic fields and microwave (MW) sources used for trapping and cooling ion chains. In this section we discuss the results of the implementation of the trap, where the characterization of the confinement capabilities of the trap are evaluated and compared with the theoretical design. This section closes reporting the optimal trap operation parameters, employed in this work, for trapping $^{174}$Yb$^+$ and $^{171}$Yb$^+$.

## II. DESIGN AND CONSTRUCTION

### A. Linear Paul trap

A Paul trap is a quadrupole trap that uses a radio frequency (RF) electric field generating an effective pseudopotential which, in combination with a constant potential, generates a minimum where ions can be trapped. These potentials are applied to electrodes, whose geometry defines the trapping potential and the capabilities of the trap.

In this work, we use a configuration of electrodes with different potentials that generates a trapping potential, where it is possible to trap plane or linear ions crystals. The trap is composed of four rods which gives a radial confinement (X and Y axes) and an endcap in the axial direction (Z axes), where RF is imposed in the radial plane and the constant potential is held in the perpendicular plane. The total potential at the center of the trap using this geometry is:

$$\Phi(x,y,z,t) = \frac{U_{dc}}{2}(\alpha_{dc}z^2 + \beta_{dc}y^2 + \gamma_{dc}x^2) \\ + \frac{U_{rf}}{2}\cos(\omega_{rf}t)(\alpha_{rf}z^2 + \beta_{rf}y^2 + \gamma_{rf}x^2) \quad (1)$$

where $U_{dc}$ is the endcaps voltage, $\omega_{rf}$ is the angular drive frequency, $U_{rf}$ is the RF drive amplitude and the $\alpha$, $\beta$ and $\gamma$ are the Laplace coefficients that are given by the trap geometry. For the previous equation an additional constraint must be taken into consideration: $\alpha_{dc} + \beta_{dc} + \gamma_{dc} = 0$ and $\alpha_{rf} + \beta_{rf} + \gamma_{rf} = 0$, this is due to the Laplace equation in free space[28] $\Delta\Phi(x,y,z) = 0$. To achieve this constraint, we chose for the solution of $-\alpha_{dc} = \beta_{dc} + \gamma_{dc}$ and $\alpha_{rf} = 0$, $\beta_{rf} = -\gamma_{rf}$. Which correlates in an ideal linear Paul trap with dynamic confinement in the $x-y$ plane and a constant confinement in the $z$ direction.

The potential from the previous equation can be achieved with different configurations of linear traps. In this work we chose a design that utilizes blades instead of rods, which have the advantage of placing the electrodes as close as possible to the ions (increasing the confinement frequency)[27]. We depart from a design from the University of Amsterdam[29], which is a blade trap where the blades that are held at the same potential, are arranged in a fork configuration. Both pairs of blades are achieved using two independent forks insulated between each other. This trap was proven to accomplish several of the requirements we desired, but it had a high capacitance between the RF blades. To improve the region of low micromotion, the shape, dimensions and separation of the blades were optimized to achieve the optimal electric field in the center of the trap. This optimization process is described in the following section where we discuss the proposed objectives and the results of the simulation process.

### B. Simulation

The trap was designed using 3D modeling software (Solidworks) and the final trap can be seen in Figure 1. Next we introduce this model in a finite elements software for its evaluation and analysis. The evaluated region is only around the center of the trap and the information we want to extract from this simulation is a rough approximation of the electric field. From the shape of the field, the values of the parameters $\alpha$, $\beta$ and $\gamma$ are obtained.

The finite element analysis provides data on the chosen mesh for the simulation. Increasing the number of nodes in the simulation improves the resolution, but it also increases the process time. Therefore, the output of this stage is an approximation of the field at the center of the trap. This rough approximation is then fitted, taking into consideration that the field follows a quadratic shape and the main coefficient is only the quadratic term of the polynomial along each of the Z, Y and X directions corresponding to the $\alpha$, $\beta$ and $\gamma$ coefficients.

DC and RF coefficients are simulated changing the potentials of the electrodes in the finite element simulation. It is important that for each simulation the other electrodes are held at a zero volt potential (ground). For the DC coefficient, both endcaps are held at a one volt potential (for normalization purposes) and the blades are grounded. In exchange for the RF coefficients, the endcaps and one pair of the RF electrodes are grounded, and the other pair is held at one volt. This simulation is for an asymmetric drive of the trap (not differential).

The objective is to achieve a value of $\alpha_{rf}$ as close as possible to zero. To achieve this condition as better as possible, the RF electrodes should be as long and as close to each other as possible. However, the blades have a finite length and their separation must allow the passage of the lasers. Also, to improve the axial confinement, the endcaps should be as flat, big and close as possible from the center of the trap, and this requires a compromise in the design parameters.

The principal trade-off relationship in the design is between the length and distance of the blades, the distance and size of the endcaps and the final shape of the blade itself. To achieve an optimal result that ensures a condition for a trapping region that permits long ions chains, which can be cooled with reasonable trapping frequencies, an iterative process of optimization was established. Different configurations where designed, simulated and then improved upon these results.

Figure 2 shows the electric field at the center of the trap—and in a small surrounding region—for three different trap configurations, and the same applied voltages. These configurations differ in blade length, aperture, and endcap separation, as detailed in Table I. In the same figure, three pairs of vertical lines are also shown. Each pair represents the approximate length that a chain of ten ions would occupy in each configuration. The distance between the lines reflects the strength of the axial confinement: a larger separation indicates a weaker confinement, which corresponds to a lower axial frequency. As a result, the ions in the chain are spaced farther apart from each other[30].

| Trap version | a | b | c |
|---|---|---|---|
| Blade length [mm] | 6 | 8 | 8 |
| Blade aperture [mm] | 2 | 2 | 1.75 |
| Endcap separation [mm] | 8 | 10 | 10 |

TABLE I. Dimensions for the traps simulated in Figure 2.

It is possible to see that the minimum electric field variation (lowest $\alpha_{rf}$) is for the longest blade and with less aperture. A tighter confinement region makes the electric field for each ion differ to its neighboring ions in a greater amount. Which implies that the micromotion experienced by each ion will be different. An effect that can be problematic for applications where an exacerbated micromotion implies an increment on the uncertainty of measurements[31]. However traps with longer, closer RF electrodes have weaker confinement in the axial direction. This can be partly compensated by increasing the axial DC voltages. Considering these limitations, and a target axial confinement in the MHz range with potentials below 1 kV, we chose the configuration labeled **a**.

The final shape of the blades and endcap can be seen in the Figure 1. Where the side and profile of the tip of the blades, and the final position and shape of the endcaps can be seen. The dimensions used for the trap are close to the ones in configuration **a**. The only change was in the angle of the base of the blades that enables a bigger endcap surface closer to the ions and a bigger hole in the center for the passage of the laser. The electric field in the center of the trap was almost the same as the **a** version.

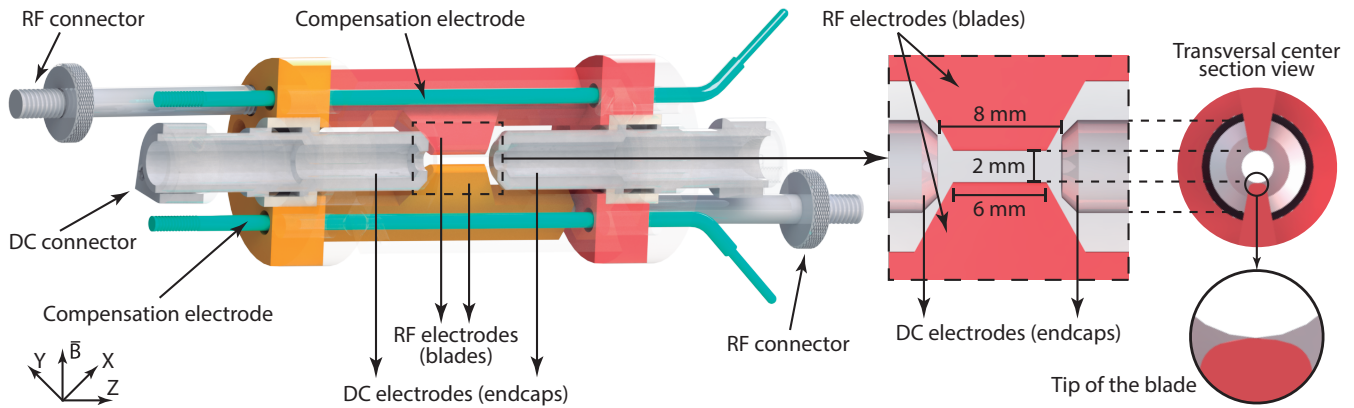

FIG. 1. 3D model of the final linear Paul trap with blade shaped RF electrodes. Half of the trap is transparent to see the center of the trap and the hollow of the endcaps to allow laser passage. A zoom of the shape of the blades can be seen on the right (square), where only one pair of electrodes is presented with the final dimensions of the trap, correspondent with the trap **a** of the table I Also it is possible to see a section view (with only two blades) of the center of the trap with a zoom on the profile of the tip of the blade.

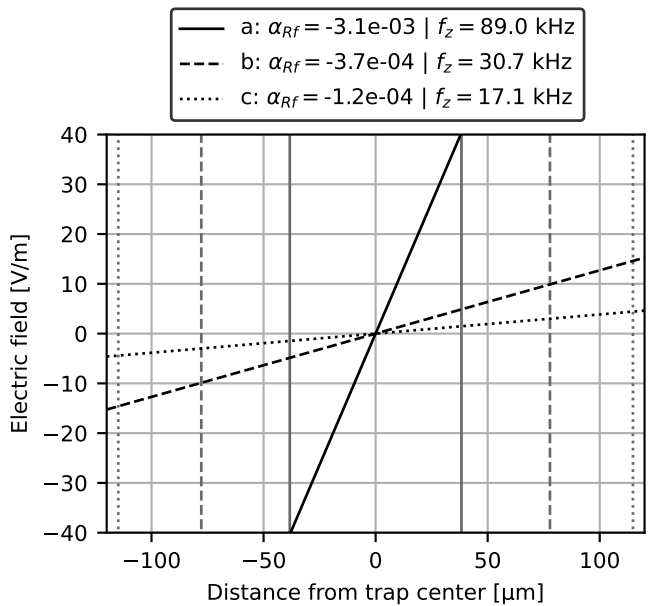

FIG. 2. The amplitude of the oscillating electric field along the axial direction $z\alpha_{rf}V$. In all cases the applied voltage is $V_{rf} = 350$ V$_p$ two opposite blades, while the others are left at ground. The vertical lines demark, the distance occupied by a crystal of ten ions in each trap, for a fixed endcap voltage of $V_{DC} = 90$ V.

### C. Excess micromotion compensation

When the potential minimum generated by the RF voltages does not coincided with the one generated by the DC voltages, excess micromotion is generated. This unwanted effect can be avoided by adjusting the DC potentials generated at the center of the trap with the use of auxiliary electrodes. We incorporated five *compensation electrodes* which can be independently controlled. Two pairs of the electrodes are in the top and bottom of the trap, as close as it was possible by design.

The fifth electrode is composed of two rods which are held at the same potential and are further from the center of the trap, to allow a better passage of the lasers. On the other side of the trap, were the detection optics is placed, no compensation electrode is placed. This increases the light collection arising from the fluorescence of the ions, as the electrodes would partly block the path to the detection system.

Another benefit of these electrodes is that they could be used to rotate 2D crystals as their distribution and geometry permit it[29,32].

### D. Construction and Assembly

In this section we discuss the design of the blade trap and the main considerations for its assembly.

The design of the insulation parts and other components was carefully optimized to minimize the capacitance, as excessive capacitance between the RF blades can strongly affect the resonant circuits used to boost the amplitude of the radiofrequency signal[33]. Building on previous designs such as the Tactica trap developed at the University of Amsterdam[29], our approach significantly reduces the capacitance by limiting the amount of dielectric material between the RF blades and ground, and by optimizing the geometry of the blade supports to increase separation and further lower the capacitance. We achieved an oveall capacity of the RF blades to all other electrodes of $\approx 40$ pF, a 4 fold reduction with respect to previous designs.

In the Figure 1 it is possible to see the shape of the blades (one blade of each pair is transparent for clarity). The endcaps are hollow to allow the passage of an axial laser. The connectors for the RF drives (knurled pieces that extrude to the sides) are screwed to the blades to improve electrical contact, and are threaded to tight a ring terminal using a nut. The endcaps are connected using ring-shaped pieces that are fastened using a worm screw that presses the connector to the cylinder. The connecting wire is tightened to the ring connec-tor using another screw which makes robust interconnection. In all cases, OFHC copper wires were used to connect the trap to the feedtroughs of the vacuum chamber.

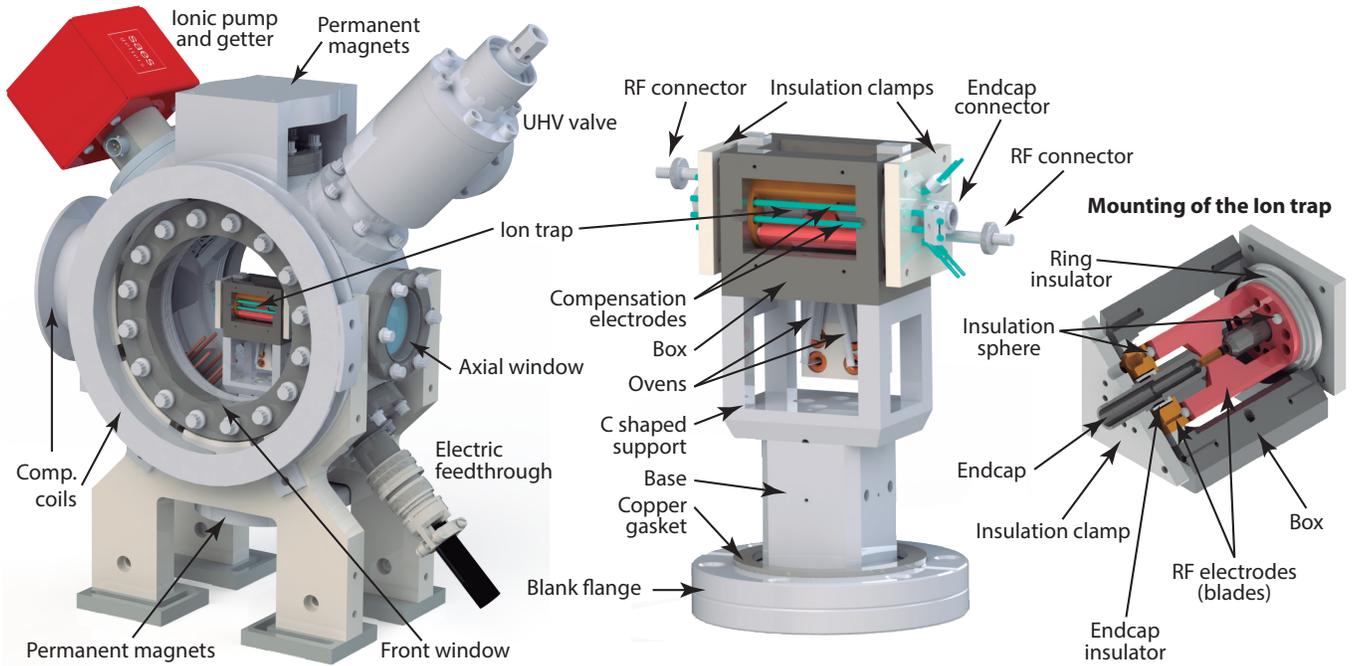

FIG. 3. 3D model of the ion trap fully mounted inside the vacuum chamber, with the outside components indicated. In the center there is a zoom of the trap support system: Box that holds the trap an allows for the insulation clamps to tight together the trap, C shaped support with apertures to improve the alignment of the ovens and base support that firmly clamps the rest of the system to the blank flange used for mounting inside the vacuum chamber. The right-most 3D model is a cut showing the assembly of the ion trap inside the box as depicted in the text.

The robustness of the trap was one of mayor considerations in the design of the blade mounting system. This system can be seen in the figure 3. It consists of three stages, the first one is the mounting *box*, which contains the blade and allows for its solid mounting. Second a *C-shaped support* which holds the ovens, and finally a *base* which sets the final height and anchors the trap to the vacuum chamber.

The mounting *box* holds the blades, separated by *insulation rings*, and allows to tightly ensemble together the blades. The RF electrodes and DC endcaps are insulated using an ring-shaped alumina piece. The blades are locked into position and insulated from each other with the use of four sapphire spheres of 2.5 mm diameter. To avoid movement of the electrodes with the vibration of the vacuum pump or just the vibrations of working on the optical table, two *insulation clamps* pull together the endcaps and hold them firmly while they push the blades to each other. The sides of the clamps allows the electrical connections of the electrodes, moving the wiring away from the passing of the lasers from the center of the trap.

The mounting *box* is attached to the a *C-shaped support* for the box, which allows the the positioning of the the ovens. We placed two ovens manufactured by AlfaVakuo using an isolated 4-terminal block. Each oven is composed of a stainless steel tube, holding granules of Yb in one case and of Ca in the other. When a current passes through the tubes, their temperature increases and emits neutral atoms. This liberation has a beam shape and the particle density is almost normally distributed[34]. The neutral atom beam must be pointed to the center of the trap, while avoiding hitting any dielectric component. Failure to do so causes the deposition of atoms that can provoke short circuits between different electrodes. Too account for this entire design was optimized to avoid direct lines of sight between the oven and the insulators. The access provided from this assembly stage is key to optimize the alignment of the ovens, which is carried out by hand pointing each oven to the center of the trap.

Finally the *base* piece holds the assembly to the vacuum chamber. It uses four vented bolts to screw it to a CF35 blank flange, which is placed on the bottom port of the vacuum chamber. The *base* part's dimensions were chosen to allow the trap to be centered in the vacuum chamber. Additionally, all other dimensions were designed to ensure that the center of the trap is axially aligned with the front and back windows. Furthermore, the endcaps are centered with the side windows to allow the passage of another laser. The vertical position of the center of the trap is 199 mm above the level of the optical table, which is less than 1 mm below the projected position, because of the copper gasket that seals the blank flange.

### E. Vacuum

The vacuum chamber is the MCF600-SphOct-F2C8 by Kimball. It has two CF100 and eight CF35 ports. The larger ports are being used for the laser entrance and for detection of fluorescense as these windows are closer to the center of the trap by design. The sides and top CF35 ports have coated windows allowing for optical access. The rest of the ports are used as follows: two for electrical access, the lower one for supporting the trap base and screw it to the rest of the chamber and the last two are one for the *ionic/getter pump* (NEXTorr Z100 by SAES group) and the other one has a VAT 57132-GE02 valve.

The valve was used to pump the whole system into ultra-high vacuum using a turbo-molecular pump-station and a carefully cleaning and baking process we describe below.

First, all trap parts, screws, wires, feedtroughs, windows and gaskets where cleaned following the notes by K. M. Birnbaum[35]. After everything was assembled, an Agilent vacuum pump system (X1699-64124) was connected to the valve, and the entire vacuum chamber was surrounded by heating elements and covered with glass wool to create an oven and keep the inside above the ambient temperature, and avoids fast heat drops that could break the windows.

The vacuum process starts with the dry pump of the vacuum system, with the valve open and using a pirani (included in the pump system) to measure the pressure. The chamber will reach a $1 \times 10^{-2}$ mbar regime in a couple of minutes, then the turbo pump can be turned on. This second pump will take the chamber below $1 \times 10^{-7}$ mbar regime. In this point we started to slowly increase the temperature of the chamber to help with the vacuum process to remove all the moister that is present inside the chamber. The pressure will increase with the temperature, but if the temperature is held constant the pressure will diminish as the turbo pump is constantly working.

The entire process of the bakeout of the chamber took twenty days. The first six days the temperature was slowly raised from 25°C to 175°C, until we noticed that the pressure was not lowering anymore (it took eleven days for this to happen), after which the temperature was lowered back to 25°C, this last stage took three days. The final pressure after this was of $1 \times 10^{-8}$ mbar.

To achieve a lower pressure we use an ionic pump with a getter element placed inside the vacuum chamber. We could not use the ionic pump during the bakeout, because of the damage that could happen to the magnets of the pump, so these where removed for the baking of the chamber. Before we finally close the valve, we flashed several times the ionic pump and activated it (the activation of the getter element was done during the bakeout) with the turbo connected, we did this until the pressure was held constant at the turning on of the ionic pump. Then we left the ionic pump turned on, and we closed the valve. During the closing action we saw a peak in the pressure that we believe was caused by a pocket of moisture and air that was placed in the back of the seal of the valve, which was fully open during the vacuum process. It is recommended to not fully open the valve for the process of bakeout. We did a second a shorter bakeout only of the valve to ensure that not moisture was left behind. The final pressure of the chamber is below $2.1 \times 10^{-11}$ mbar, which is the reading limit of the ionic pump.

### F. RF and DC drives

The trap drive consists of a series of potentials imposed on the different electrodes, to achieve the required potential (see Eq. 1).

For the RF drive a 7.2 MHz signal is imposed between one pair of the blades, and the other pair is connected to ground. The trap was usually driven with an amplitude of approximately 260 to 350 volts. To generate this, the output from a signal generator (Techtronix AFG1022) is amplified by a RF power amplifier (Mini Circuits ZHL-5W-1+), whose output voltage is raised to the desired levels using an LC tank resonator with inductive coupling[33]. This resonator was designed to resonate at the driving frequency and to have a very high Q ($195.3 \pm 0.5$). Additionally, we note that the resonator works as a selective band pass filter, reducing noise induced in the RF electrodes. The maximum amplitude voltage that can be safely supplied to the RF electrodes is 1000 volts.

The output of the resonator is referenced to ground as it is not a floating differential signal. This implies that one pair of the RF blades are held at ground potential, and the other pair is connected to the oscillating voltage. Alternatively, the trap can be driven differentially by applying oscillating voltages to both electrodes, an option we left available for future modifications.

To monitor the output of he resonator and measure the voltage applied to the blades, we use a capacitive divider composed of two coaxial wires. The first capacitor is made using a small piece of high voltage coaxial wire (RG213), where the center conductor is soldered to the output of the resonator and the capacitor is formed between this conductor and the outside metal mesh, changing the length of the piece of wire changes the capacitance as it is distributed along the wire. The second capacitor is made using a common 50 Ω (RG58) coaxial wire that is connected between the mesh of the first capacitor and the input of the oscilloscope, therefore the second capacitor is formed between the parallel capacitance of the second wire and the input capacitance of the oscilloscope. This capacitive divider was built following the guidelines of Heinz Lenk from the University of Mainz.

The resonating frequency is measured using an standing wave ratio (SWR) meter after the RF amplifier. The design of the resonator was mainly carried out following the work by Siverns et al.[33]. It is very important to have an approximate value of the capacity of the trap and a good estimation of the parasitic capacitance of the whole system. If these spurious effects are not considered, the final resonating frequency will not match the calculated one.

The DC electrodes use a continuous voltage imposed by a high voltage source (Iseg EBS C0 30 module). This is a low noise, low current tunable voltage source and can be controlled remotely through a computer network and a HTTP API. Before this source is connected to the electrodes a 1 mH inductor is placed in series to avoid the RF drive be directly induced into the Iseg channel.

The total potential in the middle of the trap are a function of the RF and DC voltages, as can be seen in the equation (1). For a given geometry, the trap have regions of stability which are mainly defined by the RF and DC potentials, as the confinement frequencies of the trap are dependent of these voltages. The stability of the trap can be defined in terms of the adimentional parameter $q$, which can be written as a function of these frequencies[36]:

$$q = \frac{2}{\Omega_{rf}} \sqrt{2\omega_r^2 + \omega_z^2} \qquad (2)$$

where $\Omega_{rf}$ is the drive frequency of the trap, $\omega_r$ and $\omega_z$ are the radial and axial confinement frequencies of the trap.

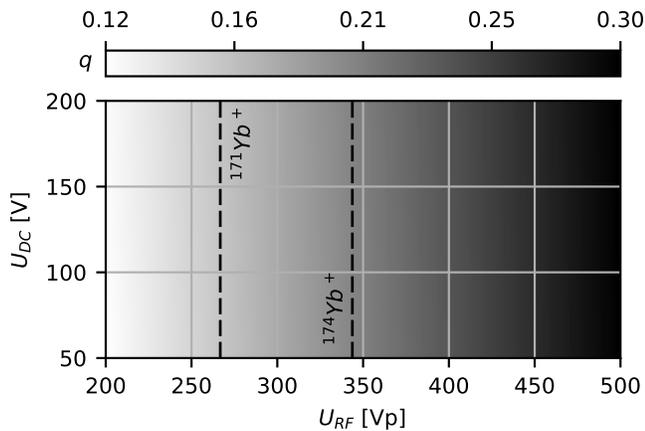

FIG. 4. $q$ parameter for the designed trap, as a function of DC and RF voltages. The values represented in vertical lines are the actual voltages used for trapping and cooling each isotope.

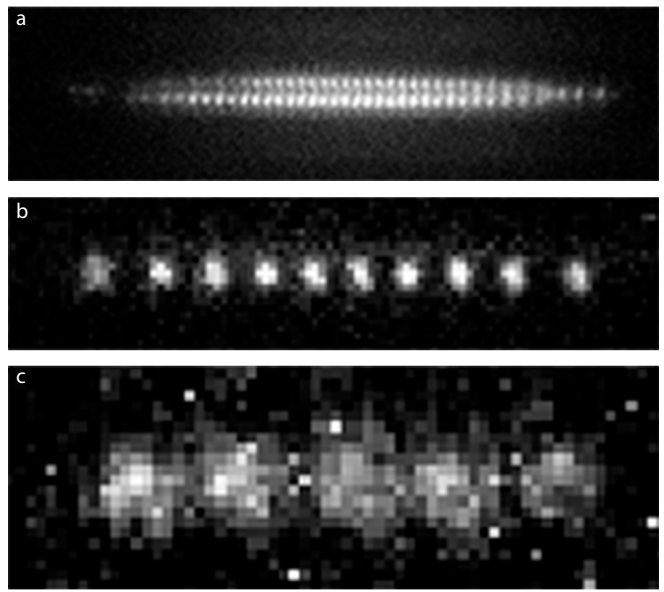

FIG. 5. Ion crystals for different Yb isotopes: a) Zeppelin shaped crystal of more than seventy ions of $^{174}$Yb$^+$. b) Linear ion crystal of $^{174}$Yb$^+$. c) Linear ion crystal of $^{171}$Yb$^+$.

In the Figure 4, the parameter $q$ of the trap is represented as a function of the DC and RF voltages and the actual voltages used for each isotope are presented as vertical dashed lines. As it is possible to see, for this configuration the $q$ lies between 0.1 and 0.3, which is a stable region as the value for $q$ in this kind of trap must lie in the range $0 < q < 0.9$[37].

## III. IMPLEMENTATION AND OPERATION

In this section we will present and discuss the rest of the elements required for trapping and carrying out experiments with ion chains as depicted in the Figure 5. This includes the calibration and characterization measurements that we compare with the simulated values.

To be able to work and experiment with ions, different instruments are required, like the detection system (camera and photomultiplier), lasers, coils, magnets and microwave electronics. These components require a fast and reliable control system to perform pulsed experiments, which allow precise control and timing of the different instruments using programmable routines.

Centralized control of the trap is performed from a computer that operates, on one hand, remotely controlled instruments with response times on the order of a second (such as laser drivers, function generators, or the aforementioned ISEG, which are controlled via USB or Ethernet). And on the other hand, through an Artiq board[38], which can generate digital and analog signals controlled in parallel with nanosecond timing precision. This platform consists of an FPGA and a micro controller, both programmable using Python language, with hardware modules composed of TTL input and output signals (Sinara 2128), digital to analog converters (Sinara 5432 "Zotino") and radio frequency generators (Sinara 4410 "Urukul"). These signals act, via acousto-optic modulators, switches, and mixers, on the laser and microwave sources, enabling rapid/real-time adjustments of frequency and power allowing synchronization of control for acquisition operations.

### A. Detection system

The detection system consists of a single 50 mm diameter aspheric lens with UV coating (Edmund Optics). The lens is placed in a mounting system that allows to adjust its angle in the X-Y plane, and three manual micro meter actuators to align each axis independently.

To improve stability and ease of configuration an *elevator* is used to lower the signal to the level of the optical table,

where two detection possibilities are placed. The first one is a sCmos camera (Andor Zyla 4.2) which allows to capture images of the ions. For small fluoresence signals, the second system is a photo-multiplier (PMT model H5783P by Hamamatsu) that generates a pulse each time a photon hits its front electrode. This pulses are analogically (custom circuit) discriminated from the dark background, which is a spontaneous generation of photons in the PMT with a lower output amplitude[39]. The differentiated pulses are amplified and normalized in amplitude to a TTL level (0 to 5 volt). These events are recorded with use of a digital input and a counter implemented in the Artiq system.

### B. Ionization of neutral Yb

As a source of Yb the corresponding oven is used. The oven is heated by the injection of a direct current. The temperature of the oven grows monotonically with respect to the current circulating through it. At higher temperatures, the flux of neutral ytterbium increases correspondingly.

The neutral beam is composed of all Yb stable isotopes (169,170,171,172,173,174 and 176). The relative abundance of each isotope will impact on the observed neutral-atom fluorescence signal[40]. This effect can be seen in the spectrum of the Figure 6. The spectrum corresponds to the first step of the two photon ionization process, where a 398.9 nm laser, takes the neutral Yb from the $^1S_0$ to the $^1P_1$. The exact wavelength for this first step is dependent of the isotope, giving the possibility of doing a selective ionization process to work with a specific isotope. That is, for the use of a specific isotope, the wavelength of the first step of the ionization process must be tuned to the correspondent peak of the figure 6.

Ionization of a single isotope requires to work with a relative low oven current and low intensity of the first ionization step. As the spectrum of the isotopes widens with the increase of temperature and speed of the particle beam, and with increasing laser power. Therefore it is highly possible to trap more than one isotope due to the broadening of the transition, which will be equal for all isotopes[41].

To achieve a selective ionization of $^{174}Yb^+$ we use an oven current of 3 A and the first step ionization in the wavelength 398.9113 nm (751.526673 THz ) with a power of 100 μW and beam diameter 110 μm. For the $^{171}Yb^+$ ionization the wavelength is 398.9108 nm (751.527573 THz), with the same parameters stated before. The laser's frequency is stabilized by applying a feedforward scheme. Based on the wavelength measurement of the, using a HighFinesse WS-7 wavemeter, we modify the wavelength by acting on the laser diode current and external cavity length (Toptica MDL DL PRO HP of 399 nm).

We performed an evaluation of the wavemeter which has an absolute accuracy, according to the manufacturer, of 600 MHz. The wavemeter reports data with a resolution below 5 MHz, however this signal fluctuates randomly over 50 MHz in a few seconds, setting a bound to the effective precision in the short term. The wavemeter was self calibrated using an internal Neon lamp reference, and the calibration of the wavemeter is currently overdue. The absolute values obtained for the transition frequencies show a offset from reported data[42,43] in the 150 MHz range, which is consistent with the absolute accuracy of the device and did not pose any inconvenience in the repeatability of the trapping.

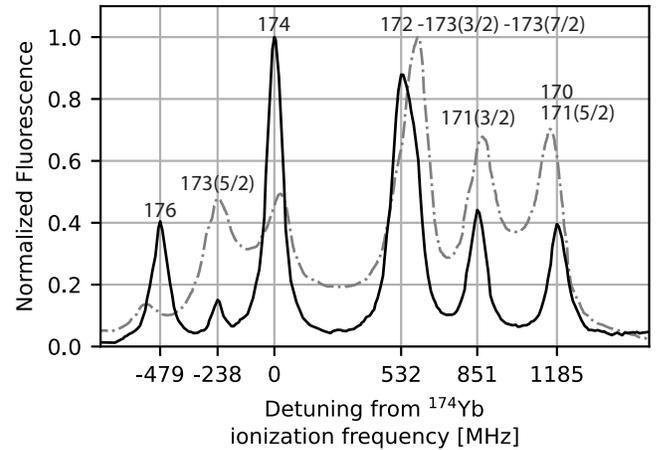

FIG. 6. Neutral spectrum of ytterbium isotopes for two different angles between the laser and the oven output. In solid black is with an angle of 95°, the dashed dotted curve is for 99°. The labels corresponds with the correspondent isotope.

By construction, the oven is not placed perpendicular to the propagation of the lasers. This slight angle is needed to allow the placement of the two ovens. The oven's angle with respect to the beam propagation direction ($\theta_\text{oven}$) gives an a Doppler shift and broadening of the spectrum, proportional to $\cos\theta_\text{oven}$. This broadening can induce the trapping of other isotopes beyond the one selected by the laser frequency. Making this angle as close to $\theta_\text{oven} = 90°$ as possible is necessary to have a selective ionization. For this reason the first step in the ionization enters trough the same point as the rest of the cooling and repump lasers, as this is the minimum angle achievable with the selected geometry. To see the difference in the spectrum entering through different angles, in the Figure 6 is it possible to see the spectrum of the isotopes of Yb entering through the front of the trap ($\theta_\text{oven} = 95°$) vs using the side windows ($\theta_\text{oven} = 99°$).

For the second step of the ionization process requires a wavelength lower than 394 nm and around 100 μW of power[44]. For this we use a 369.5 nm laser, that is the same one used to excite the $^2S_{1/2} \rightarrow {}^2P_{1/2}$ transition of the $Yb^+$ isotopes. This allows to immediately start the cooling of the ions as soon as they are ionized[45].

### C. Even isotope structure: $^{174}Yb^+$

In this section we now discuss the level scheme of $^{174}Yb^+$ and the lasers used for Doppler cooling and fluorescence imaging.

The even isotope structure lacks of hyperfine structure, therefore it is possible to achieve a closed fluorescence cycle using only two lasers, as shown in the inset of Figure 7. The main fluorescence cycle corresponds to the $^2S_{1/2} \rightarrow\ ^2P_{1/2}$, which is the fluorescence collected by the detection system. This transition is driven by the 369 nm laser, which is also used for cooling the ion crystal using a Doppler scheme[45]. The 935 nm laser is used to repump the decay to the $^3D_{3/2}$ level, which makes a closed fluorescence cycle.

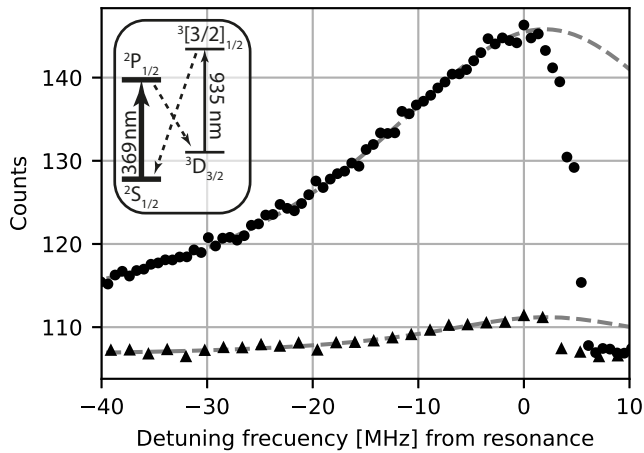

FIG. 7. Spectrum for the Doppler (369 nm, bold transition in the inset) laser with Lorentzian fit for two different intensities, the circular markers corresponds to 100 µW; the triangles depicts a 5 µW laser power. The inset represents the basic level structure for a closed fluoresence cycle of $^{174}$Yb$^+$.

The 369 nm (Toptica MDL DL pro HP) and the 935 nm (Toptica MDL DFB pro) lasers are stabilized to an external Fabry Perot (FP) cavity using a side of fringe scheme incorporated into the Toptica lasers controller. One mirror of each FP cavity was mounted in a piezoelectric transducer (PZT), which allowed us to sweep the stabilized frequency of the laser, using a slow applied voltage.

The $\gamma$ of the $^{174}$Yb$^+$ for this transition is expected to be 19.6 MHz without any broadening or heating. In the Figure 7 the spectrum of the 369 nm transition is presented for two different power (100 µW and 5 µW, measured before the trap). Each set of measurements is fitted with a Lorentzian curve, and the $\gamma$ of each is of 42 MHz (higher power) and 27 MHz (lower power), this represents the broadening due to the power of the laser of the observed spectrum. The residual broadening could be because of the presence of excess micromotion due to an incorrect compensation.

We also note that an important part to take into account for $^{174}$Yb$^+$, is the Zeeman splitting of the $^3D_{3/2}$ level, which is used for the calibration of the polarization of the lasers and it will be discussed in the III D section.

Another level to take into consideration is the $^2F_{7/2}$ level, which corresponds to an octupolar transition that can be used for frequency standards[46]. This transition, is a dark state, and can be driven by the collision of the ions with background gases[47]. Therefore another repump named 760 nm is used to drive the electronic state of the ion to the $^1[3/2]_{1/2}$ state, that will decay to the ground state. There are another ways to depopulate this state, but as was studied by J. Hur[48], using a 760 nm laser is more efficient.

To ensure that all the lasers interact with the ions is necessary to make superposition of all the beams previous to the entrance of the trap. Using a dichroic mirror allows the superposition of the Doppler, ionization and repump lasers. There are another two mirrors that allow a fine tuning of the direction of the lasers to pass through the center of the trap.

### D. Magnetic field

The magnetic axis is set to be vertical (As seen in Figure 1). To achieve this several construction and calibration considerations are in hand.

By construction two opposite rings of permanent magnets (made of Samarium-Cobalt with a diameter of 6 mm and a 10 mm height) are placed in top and bottom of the vacuum chamber, and make a magnetic field which is mainly perpendicular to the optic table. These rings are not the only magnetic fields present in the lab, the earth magnetic field and spurious magnetic fields created by wiring around the lab should be taken into consideration. To compensate this effect two extra coils are added, one to the front of the chamber and another to the side. The coils are used to adjust and correct the magnetic field generated, but it is necessary a signal to be able to define the current passing through each coil.

To callibrate the current of the coils we use a polarization dependent fluorescence signal of the trapped ions.

In our setup all lasers propagate parallel with respect to the optical table. With this orientation a vertically polarized laser is seen by the ions as a $\pi$ polarization while a horizontally polarized one is seen as $\sigma_+$ and $\sigma_-$ polarization.

We can align the direction of the magnetic field to the polarization of the 935 nm laser taking into account the following consideration. When the polarization of the 935 nm laser is parallel to the magnetic field (purely $\pi$), the ion popullation will be partly trapped in the $|^3D_{3/2}, m = \pm 3/2\rangle$ where it can only spontaneously decay to the $^2S_{1/2}$ with a lifetime of 52.7 ms. In contrast if the magnetic field is not perfectly aligned to the polarization of the 935 nm it will have some $\sigma_+$ and $\sigma_-$ component, which depopulates the D manifold and decays through the $^3[3/2]_{1/2}$ states with a decay time in the ns range. This large difference between decay times will imply that the ion will be in a dark state most of time when the the magnetic field is well aligned to the laser's polarization and in a bright one, otherwise.

To align the magnetic field we first align the polarization of the 935 nm repump to be perpendicular to the breadboard using a polarizing beam splitter. Then we adjust the current of the coils to achieve a dark ion chain, to a level where it cannot be distinguished from the background. It is important to repeat this process a couple of times during the first time of the coil activation, because the coil's wiring will change its resistance according to the temperature increase arising from the circulating current. The alignment of the polarization of the lasers respect to the magnetic field is a crucial step for trapping and cooling $^{171}$Yb$^+$.

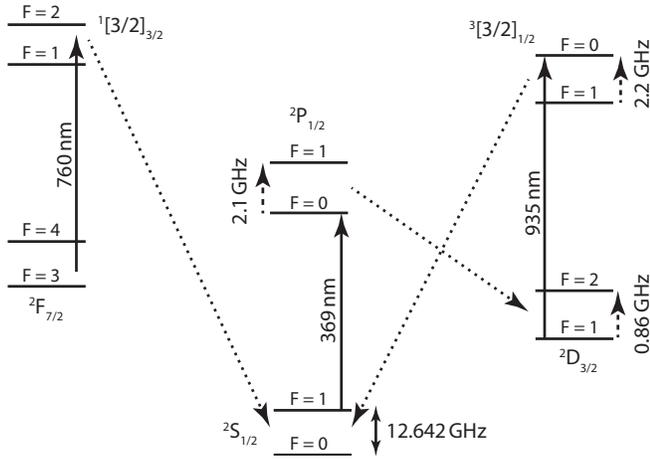

FIG. 8. Structure level diagram of $^{171}\text{Yb}^+$.

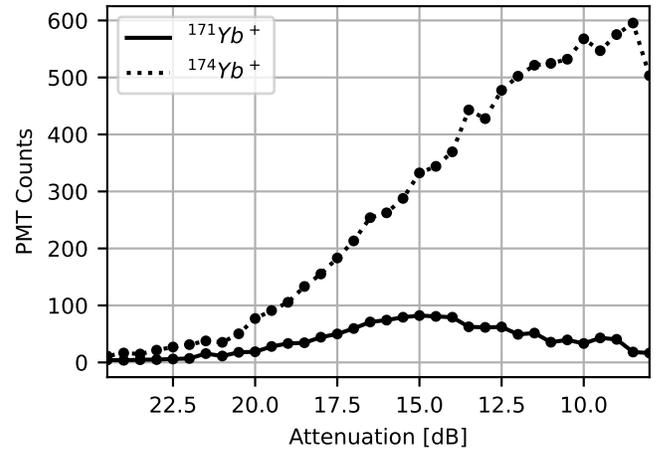

FIG. 9. Fluorescence of two different ions chains of different ytterbium isotopes.

### E. Odd isotope structure: $^{171}\text{Yb}^+$

In this section we present the level scheme of the $^{171}\text{Yb}^+$, which is similar to the previously reviewed structure of the $^{174}\text{Yb}^+$ but the main differences will be addressed.

In the Figure 8 it is possible to see the level structure of the most relevant levels of $^{171}\text{Yb}^+$ for this work. In continuous lines the transitions driven with lasers are presented. The transitions presented with dashed lines are addressed using frequency modulated lasers. Frequency modulation is achieved using using an EOMs (Qubig) which generate side bands of the desired frequency. Additionally a MW source near 12.642 GHz drives the hyperfine transition between the F states of the ground $^sS_{1/2}$ manifold.

In the case of the dipolar transition the 369 nm laser drives transitions between the ground F = 1 and the excited F = 0 state. When the laser is modulated at 2.1 GHz one can excite the $|^2P_{1/2}, F = 1\rangle$ manifold. This tranistion allows fast preparation of the F = 0 ground state, for other experiments this EOM was turned off. Even with this EOM off, the 369 nm laser off-resonatntly popullates the excited F = 1 state, which can decay to the ground F = 0 state[49]. To bring the electron back to the resonance cycle, we use the MW source which connects the ground $|^2S_{1/2}, F = 0\rangle$ and $|^2S_{1/2}, F = 1\rangle$ states.

For the repump laser (935 nm) another EOM (Qubig) is used with a drive frequency of 3.07 GHz, and using the red sideband (3.07 GHz is substracted from the frequency of the laser) the transition $|^2D_{3/2}, F = 2\rangle \rightarrow |^3[3/2]_{1/2}, F = 1\rangle$ is addressed. All the EOM where controlled using a fast switch (Minicircuits ZFSWA2-63DR+) that allows a TTL input to enable (or not) the output.

To be able to have a high degree of control of the S-P transitions, the 369 nm is separated into two different and orthogonal polarizations ($\pi$ and $\sigma_{\pm}$, respect to the magnetic field), and each path is controlled with an independent AOM. This allows to change the amount of power and to fast turn on or off $\sigma$ or $\pi$ transitions independently. Both lasers are combined into the same optic fiber to reach the ion trap. These two beams can interfere and generate unstable maxima or minima on the ion. To avoid this effect one of the AOM's drive frequency was shifted 100 kHz respect to the other.

### F. Fluorescence and cooling of $^{171}\text{Yb}^+$

There is a significant effect present in the $^{171}\text{Yb}^+$ transitions, that affects drastically the cooling process. Coherent population can generate dark states in the 369 nm dipolar transition, which strongly affects the observed fluorecence. When the exciting laser power of 369 nm increases over a certain point, dark states are stimulated[50,51] and fluorecence drops. This can be seen in the Figure 9 where the fluorescence from a four ion chain versus the attenuation of the AOM's radio frequency signal can be seen (increasing the attenuation implies lowering the laser power). The fluorescence of $^{171}\text{Yb}^+$ has a maximum and then drops. This is in contrast to the saturation behavior that $^{174}\text{Yb}^+$ presents.

The measurements of Figure 9 are normalized by substracting the background light for each point. The background was measured by turning of the repump laser. In this configuration the ions go into a dark state and most fluoresence is inhibited. Moreover the absence of the 935 nm repump laser does not modify the background counts because the the PMT has a very selective filter for 370 nm. Further, we note that the two curves presented in Figure 9 correspond to measurements taken with different numbers of ions: four $^{171}\text{Yb}^+$ ions and three $^{174}\text{Yb}^+$ ions, respectively. Additionally, the horizontal axis, which shows the relative attenuation of the laser beam, corresponds to different absolute optical powers in each case, since the incident laser power and the AOM frequencies used for the two isotopes were different. For the odd isotope, maximum fluorescence is achieved at 27 µW, with a beam diameter of 71 µm.

### G. Trapping of $^{171}$Yb$^+$

As stated in the previous sections, maximum fluorescence is achieved at 27 μW, but the second step for the ionization requires 100 μW. This difference implies that the protocol for catching and cooling $^{171}$Yb$^+$ require a change in power. This is achieved using an AOM in the 369 nm laser, to control the output power and also allows us a fast turn on and off the lasers for pulsed experiments.

To ensure reliability in the trapping process the following procedure was established:

1. Turn on the oven and slowly rise it to 3 A.

2. Turn on the MW and set the frequency to excite the most fluorescent transition.

3. Turn on the 399 and 369 nm lasers, and expose the center of the trap to both lasers at 100 μW. The amount of time depends on the size of the crystal, for a single ion about three seconds of exposition should be enough. The frequency of the 369 nm lasers must be red detuned 200 MHz, to improve the cooling process[45].

4. Lower the power of the 369 nm to 27 μW and turn off the 399 nm. Slowly move the frequency of the 369 nm to the resonant frequency of the dipolar transition.

5. Adjust the 935 nm repump if needed to ensure maximum fluorescence.

6. Finally you should have a single ion or a chain as can be seen in the Figure 5 c).

Trapping $^{174}$Yb$^+$ is similar, but without the second and fourth steps as there is no need for the MW neither to lower the power of the 369 nm. As was previously stated to cool $^{174}$Yb$^+$, it is not necessary to have the EOM of the repump turned on.

### H. State dependent Fluorecence of $^{171}$Yb$^+$

Whether the electron is in either of the two ground subspaces ($|^2S_{1/2}, F=1\rangle$ and $|^2S_{1/2}, F=0\rangle$) can be detected via state-dependent fluorescence. When the microwave is off the $|^2S_{1/2}, F=0\rangle$ state is out of the detection cycle, so it will mainly not fluoresce. Similarly, if the electron is in the $|^2S_{1/2}, F=1\rangle$ state, it will fluoresce, as it is part of the (almost) closed cycle imposed by the 369 nm and 935 nm lasers. However, off resonant excitation to the $|^2P_{1/2}, F=1\rangle$ can move the population between the $|^2S_{1/2}, F=1\rangle$ and $|^2S_{1/2}, F=0\rangle$ manifolds. This limits the total exposure time in which the fluorescence will be state dependent. In the Figure 10, we show the measurement of the fluorescence as a function of time after the MW is turned off. We see that the fluorescence decays with a time constant of $\approx 1.5$ ms after which it settles to a constant value, slightly higher than the background, which is compatible with a predominant population of the $|^2S_{1/2}, F=0\rangle$ state. Next we repeat the experiment but modulating the 396 nm laser such that the $|^2S_{1/2}, F=0\rangle$ is strongly populated. In this case we see an even faster depopulation of the bright state and a lower final fluorescence compatible with an even bigger population of the dark $|^2S_{1/2}, F=0\rangle$ state.

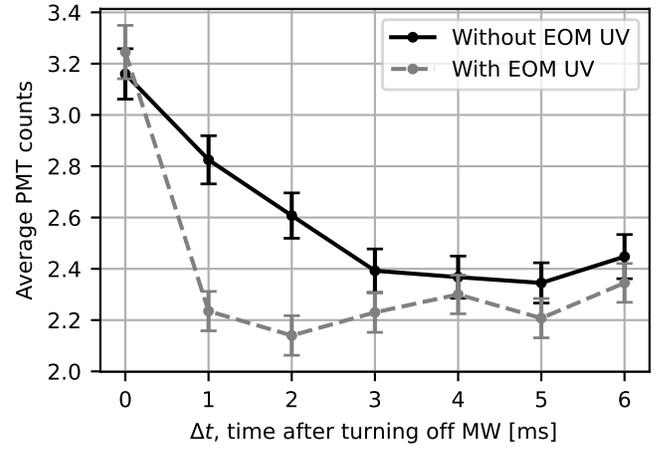

FIG. 10. Average of PMT counts function of the time after turning off the MW. It is expected that in this condition the ion decays to the ground state, therefore it won't emit any photons due to the absence of the MW. The two curves represents the use (or not) of the EOM in the UV laser that incentives the decay to the ground state, therefore using the EOM (dashed line) reduces the time it takes to prepare this state. Each point is the average of 400 measurements, and the error bar represents one standard deviation from the mean.

These results highlight two key points that will be important in the following discussion. First, state-dependent fluorescence measurements are effectively limited to integration times of approximately 1.5 ms, as longer durations do not provide additional information about the state and actually reduce the signal-to-noise ratio. Second, the use of the EOM with the UV laser proves to be an efficient method for fast preparation of the F = 0 ground state.

### I. Microwave

A microwave source is used to drive the magnetic dipole transition between the $|^2S_{1/2}, F=0\rangle$ and the $|^2S_{1/2}, F=1\rangle$ manifolds. This is needed for the cooling and fluoresence cycle, and is also used to drive coherent oscillations between these two levels.

Cooling and fluoresence of the $^{171}$Yb$^+$ depends on the correct frequency tuning of the microwave source. In figure 11 a) we show the fluorescene observed as a function opf the microwave frequency. We see three peaks corresponding to the $\Delta m = \pm 1, 0$ tranistions. Using the frequency separation of the the peaks, the magnetic field present on the ions can be approximated to be of 5.5 Gauss, based on that the Zeeman splitting for this manifold is 1.4 MHz/G.

The difference in amplitude and width of the peaks may be due to the polarization of the microwave, as the magnetic field of electromagnetic wave is mainly perpendicular to the constant magnetic field present in center of the trap, therefore

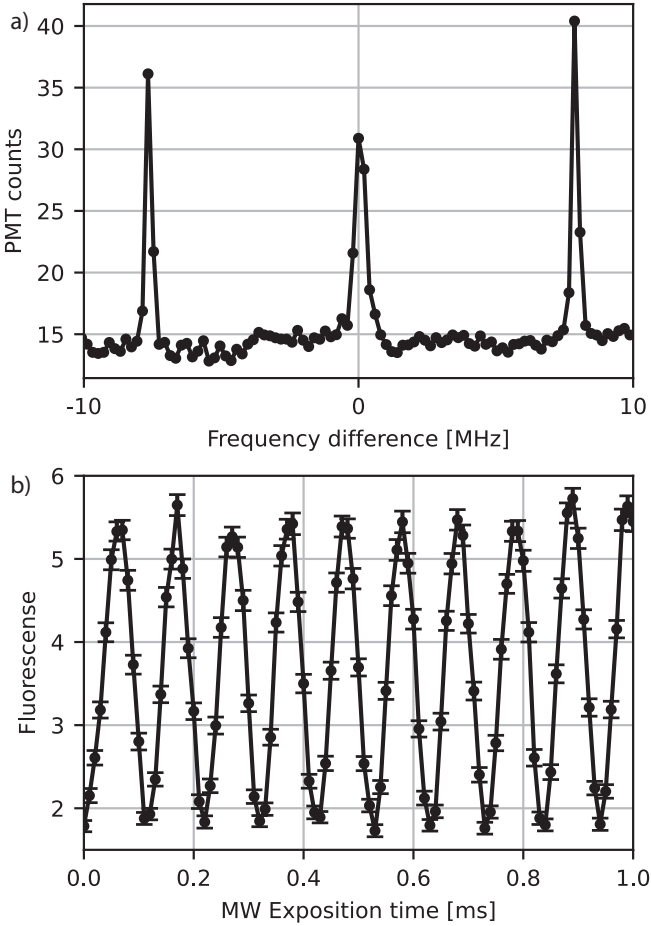

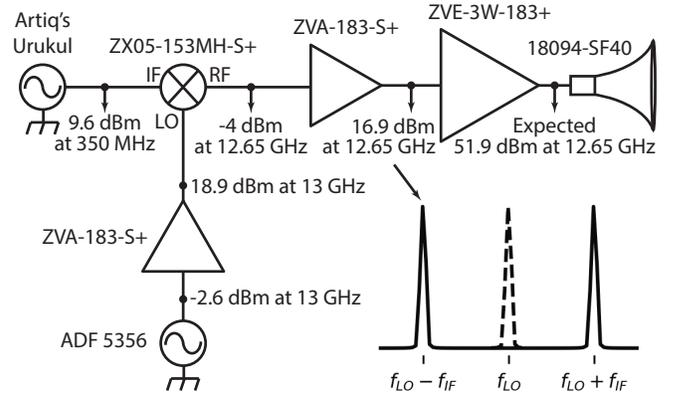

FIG. 12. Schematic of the Microwave signal, with the measured power at each stage with an inset showing a representation of the outuput of the mixer. In this case $f_{LO} = 13$ GHz and $f_{IF} = 350$ MHz as presented in the figure.

FIG. 11. a) Spectrum of the $|^2S_{1/2}, F=0\rangle \to |^2S_{1/2}, F=1\rangle$. The frequency of the transition for m = 0 (center peak) is of 12.64 GHz, and the frequency difference is ±7.76 MHz respect to each of m = ±1 (side peaks). b) a Rabi oscillation of the $|^2S_{1/2}, F=0\rangle \to |^2S_{1/2}, F=1, m=0\rangle$

the transitions involving the the m = ±1 states (which are at correspond to the peaks at the extemes) are more likely to be excited. For this reason we choose the m = 1 transition for trapping and cooling.

In the Figure 12 the schematic of the MW signal is presented. It consist in an Analog Devices ADF5356 evaluation board as a signal generator for the 13 GHz, preamplified (Mini-Circuits ZVA-183-S+) connected to the LO input of a mixer (Mini-Circuits ZX05-153MH-S+). The output amplified in two stages (Mini-Circuits ZVA-183-S+ and ZVE-3W-183+) which drive a coaxial to air adapter (Flann 18094-SF40) connected to a generic impedance matching horn for the desired wavelength. The mixer is used to create a frequency signal difference by up-conversion between the MW generator and an RF signal, that is driven by an output of an Artiq module (Urukul with AD9910) at approximately 350 MHz. This allows to fine tune the frequency emitted by the horn, and also for a fast turn on and off used for pulsed experiments. Between the evaluation board and the LO input of the mixer a preamplifier is placed, which was needed to achieve the correct power that this passive mixer requires. Using it with insufficient power can lead to unwanted non-linear behavior at it's output.

In the Figure 12, we show the signal level in between each stage. Measurements were perfomed with and spectrum analyzer (HP 8563E), taking into account the attenuation of transmission lines used (Mini-Circuits 086-36SM+ and 086-10SM+). The output power before the coaxial to air adapter, is reported as a 35 dB gain from the previous stage, as it could not be measured directly because it surpasses the maximum power input of the spectrum analyzer.

In the inset of Figure 12 we show a schematic representation of the frequency output of the signal ater the mixer. Ideally only the spectral components of $f_{LO} - f_{IF}$ and $f_{LO} + f_{IF}$, should be present[52], as it only multiplies the signals from the LO and IF ports. One of these components, represented in solid lines, is used to address the atomic transitions. We note that, as the circuit used is a ring modulator, part of the local oscillator leaks to the output, represented as a dashed line in the figure. We see a leakage of a similar amplitude as the sidebands, for the typical powers used. This has to be taken into account when chosing the frequency of the LO and IF to ensure that only one frequency component is resonant with relevant the atomic transition.

### J. Rabi oscillations

Rabi oscillations were carried out on the $|^2S_{1/2}, F=0\rangle \to |^2S_{1/2}, F=1, m=0\rangle$, the measurement can be seen in Figure 11 b).

Each point of the measurement consists in the average of 400 measured samples, with the error bars representing its standard deviation from the mean. The samples where obtained following a protocol which starts with a crystal of six ions. The steps of the measurement protocol are as follows:

1. Cooling stage: All the lasers, EOM and microwave are turned on.

2. State preparation: We prepare the ground state by turning off the microwave, everything else is turned on for 500 µs, it is basically waiting for everything to decay to the $|^2S_{1/2}, F=0\rangle$ state.

3. Exposition time: Lasers are turned off and the microwave is turned on for a variable period of time (x-axis, Figure 11 b)).

4. Detection: The lasers are turned on and the MW and the UV-EOM are off. Only if the ion is in the $|^2S_{1/2}, F=1, m=0\rangle$, it will fluoresce. The PMT counts the emitted photons during 100 µs.

5. Repetition: The previous steps are repeated 400 times for each exposition time of the microwave.

### K. Confinement measurements

In this section we will discuss three different characterizations of the trap and we will compare them with the simulation of theoretical expectations of the design.

We measured the axial and radial trap frequencies using the *tickling* method. In this technique, an additional electrode in the trap—typically an endcap or a compensation electrode—is driven with an oscillating voltage whose frequency is swept over a given range. When the driving frequency matches one of the ion's motional resonances, the ion's motion is excited, which results in a visible elongation of the ion image on the camera. This allows us to identify the resonance frequencies corresponding to the different motional modes.

In the Figure 13 b), the radial frequency of the X axes for different RF drive amplitudes can be seen. The curve fit gives $\gamma_{dc} = -0.0032$ mm$^{-2}$ and $\gamma_{rf} = 1.07$ mm$^{-2}$, which are marginally different from the original simulation ($-0.0028$ mm$^{-2}$ and $0.988$ mm$^{-2}$). These differences are believed to caused by shape of the blade. The simulations where carried out a polynomial shape, while the real blades appear to have a rounded termination. Morover, the difference in the DC coefficient in the radial direction, is suspected to be due to a slight mismatch in the axial position of the endcaps.

The slight axial mismatch of the endcaps is confirmed though the observation of the equilibrium position of the ions. We measured that as the axial confinement voltage was changed, the ion chain moved axially. To quantify this effect we varied the voltage in one endcap (L) and adjusted the voltage of other one to keep the center of the ion chain in a fixed position. We used a reference 192 V in each endcap to define the target equilibrium position. The results of these corrections can be seen in the Figure 13 a).

As an extra confirmation of the mismatch of endcap positions, we measured the axial frequency for different endcap voltages. The measurement can be seen in the Figure 13 c) together with the result of two different simulations. The original (solid), with the endcaps placed at the same distance from the center and the new simulation (dashed), that is with the endcap on the right moved 100 µm away from the center. This is consistent with the expected manufacturing tolerance of about 100 µm. We see the new simulation matches the measurement within experimental error. This gives a modification $\alpha_{DC}$ trap parameter, as the original simulation was of 0.00679 mm$^{-2}$ and the new one is 0.00709 mm$^{-2}$.

### L. Trap operation parameters

The main operation trap parameters for $^{171}$Yb$^+$ and $^{174}$Yb$^+$ are presented in this section. In the table II the RF and DC parameters are presented and the tables III to V depicts the lasers and opto electronics parameters.

| Isotope | RF drive frequency [MHz] | RF drive Amplitude [Vp] | DC Endcaps voltage |
|---|---|---|---|
| $^{174}$Yb$^+$ | 7.262 | 343.74 | 50-200 |
| $^{171}$Yb$^+$ | 7.262 | 266.76 | 50-200 |

TABLE II. Electric parameters of the trap.

| Isotope | Cooling and fluoresence (369 nm) | First Ionization step (399 nm) | Repump (935 nm) | Repump (760 nm) |
|---|---|---|---|---|
| $^{171}$Yb$^+$ | 369.5259 | 398.9108 | 935.1873 | 760.0761 |
| $^{174}$Yb$^+$ | 369.5249 | 398.9113 | 935.1792 | 760.0746 |

TABLE III. Wavelength [nm] measured using a Highfinesse WS-7 wavemeter (absolute accuracy of 600 MHz, see text), for the main transitions of $^{171}$Yb$^+$ and $^{174}$Yb$^+$. The reported wavelengths are measured previously any AOM or EOM interacting with the laser.

| Isotope | Cooling and fluoresence (369 nm) | First Ionization step (399 nm) | Repump (935 nm) | Repump (760 nm) |
|---|---|---|---|---|
| Beam diameter [µm] | 71.5 | 110 | 104 | 154 |
| $^{171}$Yb$^+$ | 27 µW | 108 µW | 1.2 mW | 2 mW |
| $^{174}$Yb$^+$ | 110 µW | 108 µW | 1.2 mW | 2 mW |

TABLE IV. Beam diameter and power of the laser used for the main transitions of $^{171}$Yb$^+$ and $^{174}$Yb$^+$.

| | EOM UV | EOM IR | Microwave $|^2S_{1/2}, F=0\rangle \to |^2S_{1/2}, F=1, m=0\rangle$ |
|---|---|---|---|
| Frequency [GHz] | 2.110 | 3.070 | 12.642817 |

TABLE V. Frequency of the EOM and MW used for $^{171}$Yb$^+$ (transitions depicted in Figure 8).

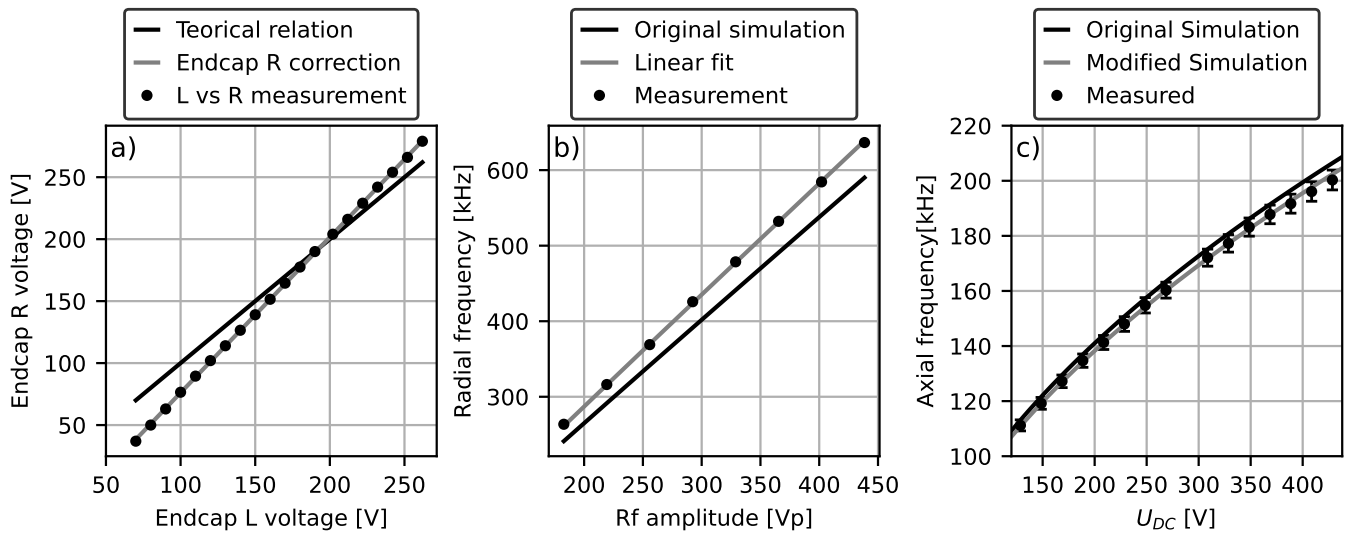

FIG. 13. Three different characterizations of the trap: a) Correction of the DC voltage in one of the endcaps due to the trap imperfection. b) Radial confinement frequency of the trap measurement as a function of the RF amplitude, measured and simulated c) Axial confinement frequency as a function of the DC voltage, measured and simulated.

## IV. CONCLUSION

A linear Paul trap was designed using 3D modeling software, simulated with finite element analysis and fully implemented in a Blade trap configuration. Cristals of $^{174}$Yb$^+$ and $^{171}$Yb$^+$ where successfully trapped and cooled. The trap was characterized using different measurements to evaluate the most important characteristics of the design.

The final result was evaluated and compared with the theoretical design, and the implementation matched closely the expected result. The deviations from the ideal trap were mainly due to the manufacturing tolerances of the required precise pieces and do not pose a problem.

Finally, pulsed experiments involving state preparation, detection, and coherent manipulation of the hyperfine levels of $^{171}$Yb$^+$ were successfully performed, demonstrating the capabilities of the ion trap and the associated laser systems, electronics, and instrumentation required for such experiments.

## ACKNOWLEDGMENTS

We gratefully acknowledge René Gerritsma for sharing the ideas and concepts underlying the original design, and Ferdinand Schmidt-Kaler for his valuable assistance in the fabrication of the trap. We also thank the undergraduate students and graduate colleagues from the LIAF for their contributions to the project. This work was financially supported by the Alexander von Humboldt Foundation (Germany); PME2015-0035; FONCyT (Argentina); the Agencia I+D+i through Grants No. PICT2016-2697, No. PICT2018-3350, No. PICT2019-4349, No. PICT2020-SERIEA-00959, and No. PICT2021-I-A-01288; Universidad de Buenos Aires through Grants No. UBACyT2018-20020170100616BA and No. UBACyT2023-20020220400119BA; and by CONICET (Argentina).